\documentclass[11pt]{article}
\usepackage[margin=1.3in]{geometry}
\usepackage{graphicx} 
\usepackage{authblk}  
\usepackage{hyperref} 
\usepackage[dvipsnames]{xcolor}
\usepackage{esvect}
\usepackage{graphicx}      
\usepackage{amsmath,amssymb, amsfonts,amsthm}
\usepackage{subcaption}
\usepackage{graphicx}
\usepackage[dvipsnames]{xcolor}
\usepackage{enumitem}
\usepackage{url}
\usepackage{caption}
\usepackage{amsmath}
\usepackage{array} 
\usepackage{amsfonts}
\usepackage{amssymb}
\usepackage{graphicx}      

\usepackage{tikz}
\usepackage{pgfplots}
\pgfplotsset{compat=1.8}
\usepackage{pgfplotstable}
\usetikzlibrary{3d, calc, decorations.pathreplacing, decorations.markings}
\usepackage{placeins}    
\pgfplotsset{compat=1.17}

\usepackage[style= numeric-comp,hyperref=true, doi=false,url=false,
            isbn=false,
            firstinits=true, sorting = none, 
            block=none, backend=bibtex,maxnames=99]{biblatex}
\renewbibmacro{in:}{} 
\bibliography{references}

\newtheorem{Definition}{Definition}
\newtheorem{Theorem}{Theorem}
\newtheorem{Lemma}{Lemma}
\newtheorem{Proposition}[Theorem]{Proposition}
\newtheorem{Corollary}[Theorem]{Corollary}

\newcommand{\cX}{\mathcal{X}}
\newcommand{\cE}{\mathcal{E}}
\newcommand{\R}{\mathbb{R}}
\newcommand{\co}{\operatorname{Co}}
\newcommand{\cH}{\mathcal{H}}
\newcommand{\bP}{\mathbf{P}}
\newcommand{\bfo}{\mathbf{1}}
\newcommand{\rint}{\operatorname{int}}

\newcommand{\xc}[1]{\vspace{.1cm}

\noindent {\em #1} }

\makeatletter
\def\blfootnote{\xdef\@thefnmark{}\@footnotetext}
\makeatother

\title{On the $H$-property for Step-graphons: The Residual Case}

\begin{document}
\author{Wanting Gao$^\dagger$\quad and \quad Xudong Chen$^\dagger$
}
\date{}
\maketitle
\blfootnote{$^\dagger$W. Gao and X. Chen are with the Electrical and Systems Engineering, Washington University in St. Louis. Emails: \texttt{\{g.wanting, cxudong\}@wustl.edu}.}
\blfootnote{This work was supported by NSF ECCS-2426017.}

\begin{abstract}               
We investigate the $H$-property for step-graphons. Specifically, we sample graphs $G_n$ on $n$ nodes from a step-graphon and evaluate the probability that $G_n$ has a node-wise Hamiltonian decomposition in the asymptotic regime as $n\to\infty$.  
It has been shown in~\cite{belabbas2023geometric,belabbas2021h} that for almost all step-graphons, this probability converges to either zero or one. We focus in this paper on the residual case where the zero-one law does not apply.
We show that the limit of the probability still exists and provide an explicit expression of it. We present a complete proof of the result and validate it through numerical studies.  


\end{abstract}

\section{Introduction}
A graphon~\cite{lovasz2006limits} $W$ is a symmetric, measurable function $W : [0,1]^2 \to [0,1]$ (i.e., $W(s, t) = W(t, s)$). It can be used as a stochastic model to sample random graphs. The sampling procedure will be described at the beginning of Subsection~\ref{ssec:Hproperty}. Graphons generalize Erd\H{o}s-R\'enyi random graph models by introducing heterogeneous edge densities for different pairs of nodes. 

The so-called $H$-property of a graphon $W$, introduced in~\cite{belabbas2023geometric,belabbas2021h}, is roughly speaking the property that a graph $G_n$ on $n$ nodes sampled from the graphon $W$ has a (node-wise) Hamiltonian decomposition almost surely as $n\to\infty$. A precise formulation will be given shortly.      
The study of the $H$-property stems from structural system theory, which investigates the impacts of networks structures (represented by graphs) on control system properties.  
The importance of having a Hamiltonian decomposition lies in the fact that it is necessary and sufficient (together with some mild condition on connectivity) for a graph to sustain ensemble controllability~\cite{chen2021sparse} and stability~\cite{belabbas2013sparse}.          

\begin{figure}[t]
    \centering
    \begin{subfigure}{.45\textwidth}
   \centering



\begin{tikzpicture}[scale=.50]

\fill[bottom color=white,top color=black] (-1.8,0) rectangle (-2.6,4) node [left] {\small$1$};
\node [left] at (-2.6,0) {\small$0$};

\filldraw [fill=black, draw=black] (0,0) rectangle (1,1);
\filldraw [fill=Gray!70!black!50, draw=Gray!70!black!50] (1,1) rectangle (2,2);
\filldraw [fill=Gray!70!black!50, draw=Gray!70!black!50] (2,2) rectangle (3,3);
\filldraw [fill=black, draw=black] (3,3) rectangle (4,4);

\filldraw [fill=black, draw=black] (3,0) rectangle (4,1);
\filldraw [fill=Gray!30!black!20, draw=Gray!30!black!20] (3,1) rectangle (4,2);

\filldraw [fill=Gray!60!black!70, draw=Gray!60!black!70] (0,2) rectangle (1,3);
\filldraw [fill=Gray!60!black!70, draw=Gray!60!black!70] (1,3) rectangle (2,4);

\filldraw [fill=Gray!30!black!20, draw=Gray!30!black!20] (2,0) rectangle (3,1);
\filldraw [fill=black, draw=black] (1,2) rectangle (2,3);

\draw [draw=black,very thick] (0,0) rectangle (4,4);

\draw[->, very thick] (0,4) -- (0,-1) node [left] {$s$};
\draw[->, very thick] (0,4) -- (5,4) node [above] {$t$};

\node [above left] at (0,4) {$0$};
\node [above] at (2,4) {$0.5$};
\node [left] at (0,0) {$1$};
\node [left] at (0,2) {$0.5$};
\node [above] at (4,4) {$1$};

\end{tikzpicture}

\caption{}
\end{subfigure} 
\begin{subfigure}{.45\textwidth}
\centering
    \begin{tikzpicture}[scale=1]
    \tikzset{every loop/.style={}}
		\node [circle,fill=cyan,inner sep=1pt,label=below left:{$u_1$}] (1) at (0, 0) {};
		\node [circle,fill=blue,inner sep=1pt,label=above left:{$u_2$}] (2) at (0, 1.75) {};
		\node [circle,fill=red,inner sep=1pt,label=above right:{$u_3$}] (3) at (1.75, 1.75) {};
		\node [circle,fill=brown,inner sep=1pt,label=below right:{$u_4$}] (4) at (1.75, 0) {};

  \path[draw,thick,shorten >=2pt,shorten <=2pt]
		 (2) edge[loop above] (2)
		 (4) edge[loop below] (4)
		 (1) edge[black] (2)
		 (1) edge[black] (4)
		 (2) edge[black] (3)
		 (3) edge[black] (4)
		 ;
\end{tikzpicture}
\caption{}
\end{subfigure}
\vspace{.3cm}

\begin{subfigure}{.45\textwidth}
\centering
    \begin{tikzpicture}[scale=1]
    \tikzset{every loop/.style={}}
		\node [circle,fill=brown,inner sep=1pt,label=left:{$v_1$}] (1) at (1, 0) {};
		\node [circle,fill=cyan,inner sep=1pt,label=left:{$v_3$}] (3) at (1, 0.9) {};
		\node [circle,fill=blue,inner sep=1pt,label=right:{$v_4$}] (4) at (2.75, 0.9) {};
		\node [circle,fill=cyan,inner sep=1pt,label=right:{$v_2$}] (2) at (2.75, 0) {};
		\node [circle,fill=blue,inner sep=1pt,label=left:{$v_5$}] (5) at (1, 1.5) {};
		\node [circle,fill=red,inner sep=1pt,label=right:{$v_6$}] (6) at (2.75, 1.5) {};

  \path[draw,thick,shorten >=2pt,shorten <=2pt]
		 (1) edge[black] (2)
		 (2) edge[black] (4)
		 (1) edge[black] (3)
		 (5) edge[black] (6)
          (3) edge[black] (4)
		 ;
\end{tikzpicture}
\caption{}
\end{subfigure}
\begin{subfigure}{.45\textwidth}
\centering
  \begin{tikzpicture}[scale=1]
    \tikzset{every loop/.style={}}
		\node [circle,fill=black,inner sep=1pt,label=left:{$v_1$}] (1) at (1, 0) {};
		\node [circle,fill=black,inner sep=1pt,label=left:{$v_3$}] (3) at (1, 0.9) {};
		\node [circle,fill=black,inner sep=1pt,label=right:{$v_4$}] (4) at (2.75, 0.9) {};
		\node [circle,fill=black,inner sep=1pt,label=right:{$v_2$}] (2) at (2.75, 0) {};
		\node [circle,fill=black,inner sep=1pt,label=left:{$v_5$}] (5) at (1, 1.5) {};
		\node [circle,fill=black,inner sep=1pt,label=right:{$v_6$}] (6) at (2.75, 1.5) {};
		
  \path[draw,thick,shorten >=2pt,shorten <=2pt]
		 (1) edge[red,bend right=12,-latex] (2)
		 (2) edge[red,bend right=12,-latex] (4)
		 (1) edge[,bend right=12,-latex] (3)
		 (5) edge[red,bend right=12,-latex] (6)
		 (3) edge[,bend right=12,-latex] (4)
         
		 (2) edge[,bend right=12,-latex] (1)
		 (4) edge[bend right=12,-latex] (2)
		 (3) edge[red,bend right=12,-latex] (1)
          (6) edge[red,bend right=12,-latex] (5)
		 (4) edge[red,bend right=12,-latex] (3)
		 ;
\end{tikzpicture}
\caption{}
\end{subfigure}

\caption{(a) A step-graphon $W$ with partition $\sigma = (0, 0.25, 0.5, 0.75, 1)$, where the value of the graphon is coded by shade, with white being 0 and black being 1. 
(b) The associated skeleton graph $S$. 
(c) An undirected graph $G$ sampled from $W$. 
(d) The directed graph $\vec{G}$ obtained from $G$, with cycles $D_1 = v_1v_2v_4v_3v_1$ and $D_2 = v_5v_6v_5$ forming a Hamiltonian decomposition of $\vec{G}$.}
    \label{fig:stepgraphon}
\end{figure}
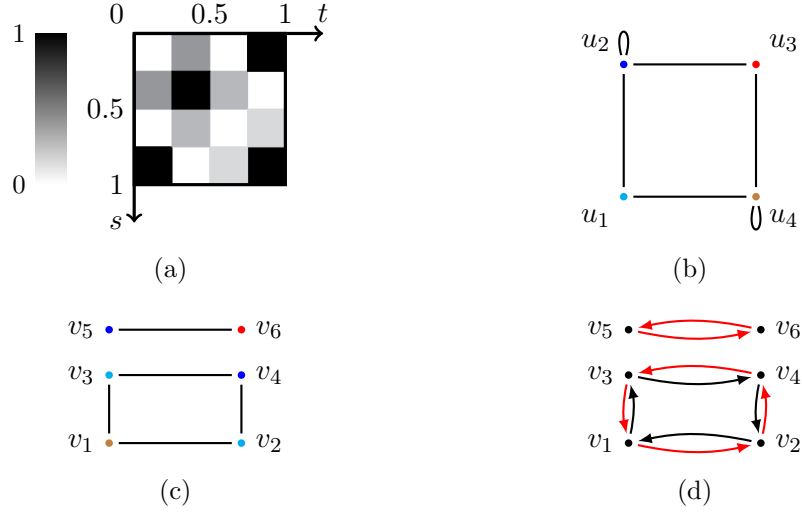

A step-graphon $W$ is a special graphon such that its domain can be partitioned into rectangles over which $W$ is constant (see Definition~\ref{def:stepgraphon} and Figure~\ref{fig:stepgraphon} for illustration). 

It has been shown in~\cite{belabbas2023geometric,belabbas2021h} that the $H$-property is essentially a zero-one property for the class of step-graphons. Specifically, for {\em almost all} step-graphons, 
the probability that the graph $G_n\sim W$ has a Hamiltonian decomposition converges to either~$0$ or~$1$ as $n\to\infty$. Moreover, whether it converges to $0$ or $1$ depends only on the support of $W$. We will review and state the result (Theorem~\ref{thm:zeroone}) in Subsection~\ref{ssec:zerooneprop}.   

In this paper, we continue to investigate the $H$-property within the class of step-graphons. We deal with what we term the ``residual case'', i.e., the case where the aforementioned probability does {\em not} converge to $0$ or $1$. Among others, we show in Theorem~\ref{thm:main} that the limit of the probability still exists as $n\to\infty$. Moreover, we provide an explicit expression of the limiting value. Our result, together with Theorem~\ref{thm:zeroone}, exhaust all possible scenarios.   

In the remainder of this section, we introduce the $H$-property and recall the almost zero-one property. Next, in Section~\ref{sec:mainresult}, we state the main result of this paper and then, in Section~\ref{sec:proof}, present the proof. In Section~\ref{sec:numericalstudy}, we conduct numerical study to validate the main result. This paper ends with conclusions. 

\subsection{The $H$-property}\label{ssec:Hproperty}
We introduce below the two-step procedure for sampling an undirected graph $G_n = (V, E)$ with $n$ nodes from a graphon $W$:
\begin{enumerate}
    \item Sample $y_1, \ldots, y_n \sim \text{Uni}[0,1]$ independently, where $\text{Uni}[0,1]$ is the uniform distribution over the interval $[0,1]$. We call $y_i$ the \emph{coordinate} of node $v_i \in V$.
    \item For any two {\em distinct} nodes $v_i$ and $v_j$, place an edge $(v_i, v_j) \in E$ with probability $W(y_i, y_j)$.
\end{enumerate}

Note that there is no self-loop in $G_n$. 

Given $G_n\sim W$, we denote by $\vec G_n = (V, \vec E)$ the \emph{directed version} of $G_n$, obtained by replacing each undirected edge $(v_i,v_j)$ of $G_n$ with two oppositely oriented edges $v_iv_j$ and $v_jv_i$. Specifically, the edge set of $\vec G_n$ is given by 
$$
{\vec{E}} := \{v_i v_j, v_j v_i \mid (v_i, v_j) \in E\}.
$$
With slight abuse of notation, we write ${\vec G_n} \sim W$. 
The directed graph $\vec{G}_n$ is said to have a (node-wise) \emph{Hamiltonian decomposition} if it contains a subgraph ${\vec{G'}_n} = (V, \vec{E'})$, with the same node set, such that $\vec {G'}_n$ is a disjoint union of directed cycles. For convenience, 
we define $\cE_n$ to be the event:
\begin{equation}\label{eq:eventforH}
\cE_n:= {\vec{G}_n}\sim W \mbox{ has a (node-wise) Hamiltonian decomposition}.
\end{equation}

The $H$-property mentioned above can now be precisely defined as follows: 

\begin{Definition}[$H$-property]\label{def:Hproperty} 
A graphon $W$ has the {\bf $H$-property} if
$$
\lim_{n \to \infty} \mathbf{P}\left ( \cE_n\right ) = 1.
$$
\end{Definition}

\subsection{Step-graphons and the almost zero-one property}\label{ssec:zerooneprop}
We start by introducing the step-graphons.

\begin{Definition}[Step-graphon and its partition]\label{def:stepgraphon}
A graphon $W$ is a {\bf step-graphon} if there exists a sequence $0 = \sigma_0 < \sigma_1 < \cdots < \sigma_q = 1$ such that $W$ is constant on each rectangle $[\sigma_i, \sigma_{i+1}) \times [\sigma_j, \sigma_{j+1})$ for all $0 \leq i,j \leq q-1$. We call 
$\sigma = (\sigma_0, \sigma_1, \ldots, \sigma_q)$ a {\bf partition} for $W$.
\end{Definition}

We next introduce the key objects that are essential for deciding whether a step-graphon has the $H$-property.

\begin{Definition}
[Concentration vector]\label{def:concentrationv} Let $W$ be a step-graphon with partition $\sigma = (\sigma_0, \ldots, \sigma_q)$. The associated {\bf concentration vector} $x^* = (x^*_1, \ldots, x^*_q)$ has entries defined as follows:
\[
x^*_i := \sigma_i - \sigma_{i-1}, \quad \text{for all } i = 1, \ldots, q.
\]
\end{Definition}

Note, in particular, that 
$$
x^*_i > 0 \quad \mbox{for all } i = 1,\ldots, q, \quad \mbox{and} \quad
\sum_{i=1}^q x^*_i = 1.
$$ 

\begin{Definition}
[Skeleton graph]\label{def:SkeletonGraph} To a step-graphon $W$ with a partition $\sigma = (\sigma_0, \ldots, \sigma_q)$, we assign the undirected graph $S = (U, F)$ on $q$ nodes, with $U = \{u_1, \ldots, u_q\}$ and edge set $F$ defined as follows: there is an edge between $u_i$ and $u_j$ if and only if $W$ is non-zero over $[\sigma_{i-1}, \sigma_i) \times [\sigma_{j-1}, \sigma_j)$. We call $S$ the {\bf skeleton graph} of $W$ for the partition $\sigma$. 
\end{Definition}

Note that the concentration vector $x^*$ and the skeleton graph $S$ combined determines completely the support of the graphon $W$. In the sequel, we assume that $S$ is connected, which is equivalent to the condition that $W$ is not block-diagonal up to a measure-preserving map on the interval $[0,1]$.

We decompose the edge set $F$ of $S$ as
$$
    F = F_0 \cup F_1,
$$
where elements of $F_0$ are self-loops, and elements of $F_1$ are edges between distinct nodes. We introduce the node-edge incidence matrix associated with $S$ in the following definition.

\begin{Definition}
[Incidence matrix]\label{def:incidencematrix} Let $S = (U, F)$ be a skeleton graph. Given an arbitrary ordering of its edges and self-loops, we let $Z = [z_{ij}]$ be the associated {\bf incidence matrix}, defined as the $|U| \times |F|$ matrix with entries:
\[
z_{ij} := \frac{1}{2}
\begin{cases} 
    2, & \text{if } f_j \in F_0 \text{ is a loop on node } u_i, \\ 
    1, & \text{if node } u_i \text{ is incident to } f_j \in F_1, \\ 
    0, & \text{otherwise.}
\end{cases} 
\]
\end{Definition}
Note that the columns of $Z$ are probability vectors, i.e., all entries are nonnegative and sum to one.

The so-called edge polytope of $S$ is defined as the convex hull of the columns of the matrix $Z$. Precisely, we have 

\begin{Definition}
[Edge polytope]\label{def:edgepolytope} Let $S = (U, F)$ be a skeleton graph and $Z$ be the associated incidence matrix, with $z_j$, for $1 \leq j \leq |F|$, the columns of $Z$. The {\bf edge polytope} of $S$, denoted by $\mathcal{X}(S)$, is the finitely generated convex hull:
\[
\mathcal{X}(S) := \left\{ \sum_{j = 1}^{|F|} c_j z_j \mid \sum_{j = 1}^{|F|} c_j = 1\mbox{ and } c_j \geq 0 \mbox{ for all } j \right \}.
\]
\end{Definition}

The rank of the polytope $\mathcal{X}(S)$ is the dimension of its relative interior, which we denote by $\rint \mathcal{X}(S)$.  
Since $\mathcal{X}(S)$ is contained in the standard $(q-1)$-simplex, its rank is bounded above by $(q- 1)$, and it can achieve full rank $(q - 1)$ if and only if the incidence matrix $Z$ has full (row) rank (i.e., rank $q$).

The following result~\cite{ohsugi1998normal} relates full rankness of $\mathcal{X}(S)$ to the existence of an odd cycle in $S$---a cycle in $S$ is \emph{odd} 
if it contains an odd number of distinct nodes (a self-loop is an odd cycle).

\begin{Lemma}\label{lem:fullrankness}
Let $S$ be the skeleton graph of $W$ on $q$ nodes. Suppose that $S$ is connected; then, 
\[
\text{rank} \, \mathcal{X}(S) = 
\begin{cases} 
q - 1 & \text{if } S \text{ has an odd cycle,} \\
q - 2 & \text{otherwise.}
\end{cases}
\]

\end{Lemma}

As mentioned earlier, it has been shown in~\cite{belabbas2023geometric,belabbas2021h} that the $H$-property is essentially a zero-one property in a sense that the probability of the event $\cE_n$ tends to be either $0$ or $1$. Precisely, we have the following result: 

\begin{Theorem}\label{thm:zeroone}
Let $W$ be a step-graphon, with $\sigma$ a partition. 
Let $x^*$ and $S$ be the associated concentration vector and the skeleton graph 
(which is assumed to be connected), respectively. 
Then, the following hold: 
\begin{enumerate}
\item If $S$ has an odd cycle and if $x^*\in \rint \mathcal{X}(S)$, then 
$$
\lim_{n\to\infty}\mathbf{P}(\cE_n) = 1.
$$
\item If $S$ does not have an odd cycle or if $x^*\not\in \mathcal{X}(S)$, then
$$
\lim_{n\to\infty}\mathbf{P}(\cE_n) = 0.
$$
\end{enumerate}
\end{Theorem}

\section{Main Result for the Residual Case}\label{sec:mainresult}
In this section, we evaluate the limit of $\mathbf{P}(\cE_n)$ as $n\to\infty$ for the case left out by Theorem~\ref{thm:zeroone}, namely, the case where $S$ has an odd cycle (so $\cX(S)$ has full rank by Lemma~\ref{lem:fullrankness}) and $x^*$ belongs to the boundary of the edge polytope, i.e.,  $$x^*\in \partial \cX(S) := \mathcal{X}(S) - \rint \mathcal{X}(S).$$

We need a few preliminaries to state the main result of this paper. 
Given a graph $G_n\sim W$, we let $n_i(G_n)$ be the number of nodes $v_j$ of $G_n$ whose coordinates $y_j \in [\sigma_{i-1},\sigma_i)$.  
Then, the {\bf empirical concentration vector} of $G_n$ is defined to be:
\begin{equation}\label{eq:empiricalcv}
x(G_n) := \frac{1}{n} \big(n_1(G_n), \ldots, n_q(G_n)\big).
\end{equation}
It follows from the step~1 of the sampling procedure in Subsection~\ref{ssec:Hproperty} that $nx(G_n)$ is a multinomial random variable with $n$ trials, $q$ events, and $x^*_i$'s the event probabilities. 

Next, we define the random variable 
\begin{equation}\label{eq:defomegan}
    \omega(G_n):= \sqrt{n}(x(G_n) - x^*) + x^*,
\end{equation}
The following result is known~\cite{belabbas2021h}:

\begin{Lemma}\label{lem:limitdist}
    The random variable $\omega(G_n)$ defined in~\eqref{eq:defomegan} converges in distribution to the Gaussian random variable $$\omega^* \sim N(x^*, \Sigma), \quad \mbox{with } \Sigma:= \operatorname{Diag}(x^*) - x^* {x^*}^\top,$$ 
    where $\operatorname{Diag}(x^*)$ is the diagonal matrix whose $ii$th entry is~$x^*_i$. 
\end{Lemma}

Let $H_0$ be the hyperplane in $\R^q$ that contains the standard simplex, i.e., 
$$
H_0 := \left \{ v \in \R^q \mid v^\top \mathbf{1} = 1 \right \}.
$$
The covariance matrix $\Sigma$ given in Lemma~\ref{lem:limitdist} has rank $(q - 1)$, with the null space spanned by the vector $\mathbf{1}$. 
In particular, since the mean of~$\omega^*$ is $x^*$, which belongs to $H_0$, the support of~$\omega^*$ is~$H_0$.  

We introduce below a convex subset of $H_0$, which will play a central role in expressing the limit of $\mathbf{P}(\cE_n)$. 
To this end, consider the polyhedral cone $\co(S)$ generated by the $z_j$'s, i.e., the column vectors of the incidence matrix $Z$ associated with $S$ (see Definition~\ref{def:incidencematrix}):
\begin{equation}\label{eq:conevertexrep}
\co(S):= \left\{ \sum_{j =1}^{|F|} c_j z_j \mid c_j \geq 0 \mbox{ for all } j\right \}.
\end{equation}
It should be clear that for any nonzero vector~$v\in \co(S)$, there exists a unique $x\in \cX(S)$ and a unique non-negative real number $c$ such that $v = cx$. 

By our hypothesis, $S$ has an odd cycle, so $\cX(S)$ has rank $(q - 1)$ (equivalently, the incidence matrix $Z$ has full rank) and hence, $\co(S)$ has rank~$q$.    

\begin{Definition}\label{def:facet}
A {\bf facet-defining hyperplane} $H_\ell$ of the polyhedral cone $\co(S)$ is a co-dimensional one subspace of $\R^q$ that satisfies the following two conditions:
\begin{enumerate}
\item There exist $(q-1)$ linearly independent vectors $z_{\ell_1},\ldots, z_{\ell_{q-1}}$ out of the columns of the incidence matrix $Z$ such that $H_\ell$ is spanned by these vectors. 
\item The subspace $H_\ell$ is a supporting hyperplane for $\co(S)$, i.e., there exists a vector $v_\ell\in \R^q$ of unit length, perpendicular to $H_\ell$, such that 
\begin{equation}\label{eq:supplane}
v_\ell^\top x  \geq 0, \quad \mbox{for all } x\in \co(S).
\end{equation}
\end{enumerate}
\end{Definition}

It should be clear that if the vector $v_\ell$ in the above definition exists, then it is unique. 
We denote by $$\mathcal{H} := \{H_1,\ldots, H_k\}$$ the collection of all facets-defining hyperplanes of $\co(S)$ and, correspondingly, $v_1,\ldots, v_k$ the associated unit vectors. 
One can thus use the so-called {\em half-space representation~\cite{motzkin1953double}} to describe $\co(S)$ (the description given in~\eqref{eq:conevertexrep} is known as the {\em vertex-representation}):
\begin{equation}\label{eq:halfrep}
\co(S) = \left \{x\in \R^q \mid v_\ell^\top x \geq 0 \mbox{ for } 1\leq \ell \leq k  \right \}. 
\end{equation}
Given the generators $z_1,\ldots, z_{|F|}$ in~\eqref{eq:conevertexrep}, one can use, e.g., Quickhull algorithm~\cite{barber1996quickhull} to find all the facets-defining hyperplanes.  

Using the half-space representation~\eqref{eq:halfrep}, we can express the edge polytope 
$\cX(S)$ as follows: 
\begin{equation}\label{eq:hrepfep}
\cX(S) = \left \{x\in H_0 \mid v_\ell^\top x \geq 0 \mbox{ for all } 1\leq \ell \leq k \right \},
\end{equation}
and, consequently, the interior $\rint \cX(S)$ and the boundary $\partial \cX(S)$ as
$$
\begin{aligned}
& \rint \cX(S) = \left \{x\in H_0 \mid v_\ell^\top x > 0 \mbox{ for all } \ell = 1,\ldots, k \right \}, \\
& \partial \cX(S) = \left \{x\in \cX(S) \mid v_\ell^\top x = 0 \mbox{ for some } \ell = 1,\ldots, k \right \}.
\end{aligned}
$$

Now, given the concentration vector $x^*$, we let
\begin{equation}\label{eq:defcH}
\mathcal{H}(x^*) = \left \{ H_\ell \in \mathcal{H} \mid  v_\ell^\top x^* = 0\right  \}.
\end{equation}
It should be clear that $x^*\in \partial \cX(S)$ if and only if $\mathcal{H}(x^*)$ is nonempty. Further, we let
\begin{equation}\label{eq:defOmega}
\Omega(x^*):= \left\{ \omega\in H_0 \mid v^\top_\ell \omega > 0 \mbox{ for all } H_\ell\in \mathcal{H}(x^*)  \right \}.
\end{equation}

With the preliminaries above, we can now state the main result of this paper:

\begin{Theorem}[The residual case]\label{thm:main}
Let $W$ be a step-graphon, and let $x^*$ and $\cX(S)$ be the associated concentration vector and the edge-polytope introduced in Definitions~\ref{def:concentrationv} and~\ref{def:edgepolytope}, respectively. If $S$ has an odd cycle and if $x^*\in \partial\cX(S)$, then
\begin{equation}
\lim_{n\to\infty} \bP(\cE_n) = \bP(\omega^* \in \Omega(x^*)),
\end{equation}
where the event $\cE_n$ is given in~\eqref{eq:eventforH}, $\omega^*\sim N(x^*, \Sigma)$ is the Gaussian random variable introduced in Lemma~\ref{lem:limitdist}, and $\Omega(x^*)$ is given in~\eqref{eq:defOmega}.
\end{Theorem}

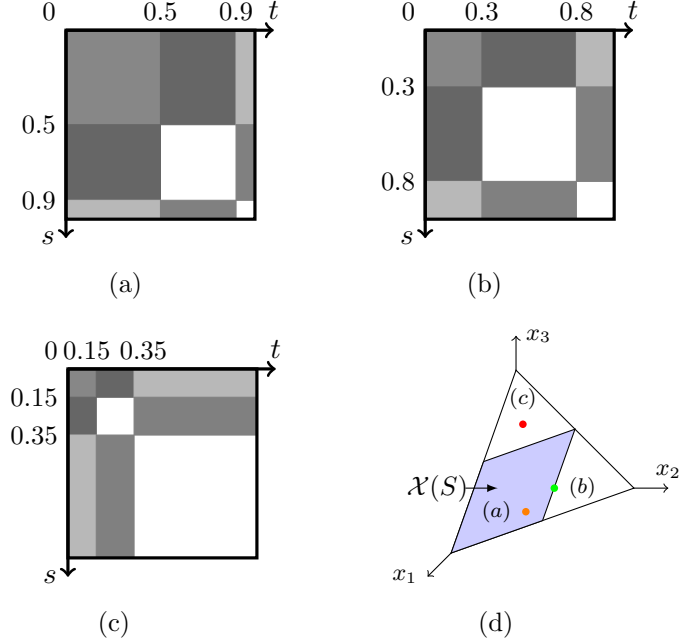
\begin{figure}[h]
    \centering
    \begin{subfigure}{0.20\textwidth}
    \centering
    \begin{tikzpicture}[scale=2.5]

        Define the partition points
        \def\xOne{0}
        \def\xTwo{0.5}
        \def\xThree{0.1}
        \def\xFour{1}

        \filldraw [fill=black!47, draw=black!47] (0,1) rectangle (0.5, 0.5); 

        \filldraw [fill=black!0, draw=black!0] (0.5, 0.5) rectangle (0.9, 0.1);

        \filldraw [fill=black!0, draw=black!0] (0.9, 0.1) rectangle (1, 0);
        
        \filldraw [fill=black!60, draw=black!60] (0, 0.5) rectangle (0.5, 0.1);

        \filldraw [fill=black!60, draw=black!60] (0.5, 1) rectangle (0.9, 0.5);

        \filldraw [fill=black!28, draw=black!28] (0, 0.1) rectangle (0.5, 0);

        \filldraw [fill=black!28, draw=black!28] (0.9, 1) rectangle (1, 0.5);

        \filldraw [fill=black!50, draw=black!50] (0.5, 0.1) rectangle (0.9, 0);

        \filldraw [fill=black!50, draw=black!50] (0.9, 0.5) rectangle (1, 0.1);

        \draw [draw=black, very thick] (\xOne,\xOne) rectangle (\xFour,\xFour);

        \draw[->, very thick] (\xOne, \xFour) -- (\xOne, \xOne - 0.1) node [left] {$s$};
        \draw[->, very thick] (\xOne, \xFour) -- (\xFour + 0.1, \xFour) node [above] {$t$};

        \node [above left] at (0, 1) {\small $0$}; 
        \node [left] at (0, 0.5) {\small $0.5$};  
        \node [left] at (0, 0.1) {\small $0.9$};  

        \node [above] at (0.5, 1) {\small $0.5$};
        \node [above] at (0.9, 1) {\small $0.9$};

    \end{tikzpicture}
    \caption{}
    \end{subfigure}
    \hspace{0.1\textwidth} 
    \begin{subfigure}{0.20\textwidth}
        \centering
            \begin{tikzpicture}[scale=2.5]

        \filldraw [fill=black!47, draw=black!47] (0,1) rectangle (0.3, 0.7); 

        \filldraw [fill=black!0, draw=black!0] (0.3, 0.7) rectangle (0.8, 0.2);

        \filldraw [fill=black!0, draw=black!0] (0.8, 0.2) rectangle (1, 0);
        
        \filldraw [fill=black!60, draw=black!60] (0, 0.7) rectangle (0.3, 0.2);

        \filldraw [fill=black!60, draw=black!60] (0.3, 1) rectangle (0.8, 0.7);

        \filldraw [fill=black!28, draw=black!28] (0, 0.2) rectangle (0.3, 0);

        \filldraw [fill=black!28, draw=black!28] (0.8, 1) rectangle (1, 0.7);

        \filldraw [fill=black!50, draw=black!50] (0.3, 0.2) rectangle (0.8, 0);

        \filldraw [fill=black!50, draw=black!50] (0.8, 0.7) rectangle (1, 0.2);

        \draw [draw=black, very thick] (0,0) rectangle (1,1);

        \draw[->, very thick] (0, 1) -- (0, -0.1) node [left] {$s$};
        \draw[->, very thick] (0, 1) -- (1.1, 1) node [above] {$t$};

        \node [above left] at (0, 1) {\small $0$}; 
        \node [left] at (0, 0.7) {\small $0.3$};  
        \node [left] at (0, 0.2) {\small $0.8$};  

        \node [above] at (0.3, 1) {\small $0.3$};
        \node [above] at (0.8, 1) {\small $0.8$};

    \end{tikzpicture}
    \caption{}
\end{subfigure}

\vspace{0.2cm} 
    
    \begin{subfigure}{0.20\textwidth}
        \centering
       \begin{tikzpicture}[scale=2.5]

        \filldraw [fill=black!47, draw=black!47] (0,1) rectangle (0.15, 0.85); 

        \filldraw [fill=black!0, draw=black!0] (0.15, 0.85) rectangle (0.35, 0.65);

        \filldraw [fill=black!0, draw=black!0] (0.35, 0.65) rectangle (1, 0);
        
        \filldraw [fill=black!60, draw=black!60] (0, 0.85) rectangle (0.15, 0.65);

        \filldraw [fill=black!60, draw=black!60] (0.15, 1) rectangle (0.35, 0.85);

        \filldraw [fill=black!28, draw=black!28] (0, 0.65) rectangle (0.15, 0);

        \filldraw [fill=black!28, draw=black!28] (0.35, 1) rectangle (1, 0.85);

        \filldraw [fill=black!50, draw=black!50] (0.15, 0.65) rectangle (0.35, 0);

        \filldraw [fill=black!50, draw=black!50] (0.35, 0.85) rectangle (1, 0.65);

        \draw [draw=black, very thick] (0,0) rectangle (1,1);

        \draw[->, very thick] (0, 1) -- (0, -0.1) node [left] {$s$};
        \draw[->, very thick] (0, 1) -- (1.1, 1) node [above] {$t$};

        \node [above left] at (0, 1) {\small $0$}; 
        \node [left] at (0, 0.85) {\small $0.15$};  
        \node [left] at (0, 0.65) {\small $0.35$};  

        \node [above] at (0.10, 1) {\small $0.15$};
        \node [above] at (0.40, 1) {\small $0.35$};

    \end{tikzpicture}
        \caption{}
    \end{subfigure}
    \hspace{0.12\textwidth} 
    \begin{subfigure}{0.2\textwidth}
        \centering
         \begin{tikzpicture}[scale=1.3]
    \tikzset{every loop/.style={}}
    \node [circle,fill=black,inner sep=0.8pt] (0) at (0,0) {};
    \draw[->] (0,0) -- (1.55,0) node [above] {\footnotesize $x_2$};
    \draw[->] (0,0) -- (0,1.55) node [right] {\footnotesize $x_3$};
    \draw[->] (0,0) -- (-0.9,-0.9) node [left] {\footnotesize $x_1$};

    \node (2) at (1.2, 0) {};
    \node (3) at (0, 1.2) {};
    \node (1) at (-0.66, -0.66) {};
    \node (12) at (0.27, -0.33) {};
    \node (13) at (-0.33, 0.27) {};
    \node (23) at (0.6, 0.6) {};

\filldraw[draw=black, fill=gray!0]  (1.center) -- (2.center) -- (3.center) -- (1.center) -- cycle;

\filldraw[draw=black, fill=blue!20]  (1.center) -- (12.center) -- (23.center) -- (13.center) -- cycle;

\node [] (xs) at (-0.8,0.0) {\small $\cX(S)$};
\node [] (xsl) at (-0.7,0.0) {};
\node [] (xst) at (0, 0.0) {};

\path[draw,shorten >=3pt,shorten <= 3pt]
    (xsl) edge[-latex] (xst);

\node [circle,fill=orange,inner sep=1.0pt,label=left:{\scriptsize$(a)$}] (x) at (0.1, -0.24) {};

\node [circle,fill=green,inner sep=1.0pt,label=right:{\scriptsize$(b)$}] (x) at (0.39, 0) {};

\node [circle,fill=red,inner sep=1.0pt,label=above:{\scriptsize$(c)$}] (x) at (0.07, 0.65) {};

\end{tikzpicture}
    \caption{}
    \end{subfigure}
    \caption{Three step-graphons shown in (a), (b), and (c) with the same skeleton graph, but different concentration vectors. The shaded region in (d) is the edge polytope $\cX(S)$, embedded in the standard simplex $\Delta^{2}$. The red, green, and orange dots are the concentration vectors $x^*$ for the above three step-graphons. We have that (a) $x^* \in \rint \mathcal{X}(S)$ and $\lim_{n\to\infty}\bP(\cE_n) = 1$; (b) $x^* \in \partial(\mathcal{X}(S))$ and $\lim_{n\to\infty}\bP(\cE_n) = 0.5$; and (c) $x^* \notin \mathcal{X}(S)$ and $\lim_{n\to\infty}\bP(\cE_n) = 0$.}
    \label{fig:2by2}
\end{figure}

Note that if $x^*$ belongs to only one facet-defining hyperplane, then $\Omega(x^*)$ is reduced to an open half-space of $H_0$, with $x^*$ on the boundary of $\Omega(x^*)$. Since the mean of $\omega^*$ is $x^*$ and since the support of $\omega^*$ is the entire $H_0$, we have that  
$\bP(\omega^*\in \Omega(x^*)) = 0.5$. 
This statement can be strengthened slightly as follows:

\begin{Corollary}\label{cor:onehalf}
    Under the hypothesis of Theorem~\ref{thm:main}, we have that
    $$
    \lim_{n\to\infty} \bP(\cE_n) \leq 0.5,
    $$
    and the equality holds if and only if $\cH(x^*)$ is a singleton. 
\end{Corollary}

\section{Proof of Theorem~\ref{thm:main}}\label{sec:proof}
Consider the events $\mathcal{A} := \left\{ x(G_n) \in \rint \mathcal{X}(S) \right\}$ and $\mathcal{B}_\varepsilon := \left\{ x(G_n)  \in B_\varepsilon(x^*) \right\}$, where $B_\varepsilon(x^*)$ is the open $\varepsilon$-ball centered at~$x^*$, and $\mathcal{A}_\varepsilon^* := \mathcal{A} \cap \mathcal{B}_\varepsilon$.
The following result can be obtained from~\cite{belabbas2023geometric}:
\begin{Lemma}\label{lem:conditionprob}
If $S$ has an odd cycle, then there exists a sufficiently small $\varepsilon > 0$  such that 
$
\lim_{n \to \infty} \bP(\cE_n \mid \mathcal{A}_\varepsilon^*) = 1.
$
\end{Lemma}
To proceed, we write 
\begin{equation}\label{eq:lawoftotalprob}
\bP(\cE_n) = \bP(\cE_n \cap \mathcal{A}_\varepsilon^*) + \bP(\cE_n \cap \neg\mathcal{A}_\varepsilon^*).
\end{equation}
We evaluate the two terms on the right hand side of~\eqref{eq:lawoftotalprob}.  

{\em First term:} We have $\bP(\cE_n \cap \mathcal{A}_\varepsilon^*) = \bP(\cE_n \mid \mathcal{A}_\varepsilon^*)\bP(\mathcal{A}_\varepsilon^*)$. 
By Lemma~\ref{lem:conditionprob}, $\bP(\cE_n \mid \mathcal{A}_\varepsilon^*)$ converges to $1$ as $n\to\infty$. For $\bP(\mathcal{A}_\varepsilon^*)$, we note that 
$$\bP(\mathcal{A}) - \bP(\neg \mathcal{B}_\varepsilon) \leq \bP(\mathcal{A}_\varepsilon^*) \leq \bP(\mathcal{A}).$$

Since $x(G_n) \to x^*$ as $n\to\infty$, $\bP(\neg \mathcal{B}_\varepsilon) \to 0$. 
Using the squeeze theorem, we obtain that $\bP(\mathcal{A}_\varepsilon^*) \to \bP(\mathcal{A})$ as $n\to\infty$. Thus, $\bP(\cE_n \cap \mathcal{A}_\varepsilon^*)$ converges to $\bP(\mathcal{A})$ as $n\to\infty$.

{\em Second term:}
We first use the fact that $\neg \mathcal{A}_\varepsilon^* = \neg \mathcal{A} \cup \neg \mathcal{B}_\varepsilon$ to obtain that
\begin{equation}\label{eq:sedterm}
\bP(\cE_n \cap \neg \mathcal{A}_\varepsilon^*) \leq \bP(\cE_n \cap \neg \mathcal{A}) + \bP(\cE_n \cap \neg \mathcal{B}_\varepsilon).
\end{equation}

From~\cite{belabbas2021h}, $\mathcal{A}$ is necessary for $\cE_n$, i.e., $\cE_n \subseteq \mathcal{A}$. It follows that $\bP(\cE_n \cap \neg \mathcal{A}) = 0$. 

Using again the fact that $\bP(\neg \mathcal{B}_\varepsilon) \to 0$, we have that $\bP(\cE_n \cap \neg \mathcal{B}_\varepsilon)$ vanishes as $n\to\infty$. Thus, $\bP(\cE_n \cap \neg\mathcal{A}_\varepsilon^*)$ vanishes as $n\to\infty$. 

Combining the arguments above, we have that $\bP(\cE_n) \to \bP(\mathcal{A})$ as $n\to\infty$. It now suffices to establish 

\begin{Proposition}\label{prop:convofprobforcv}
Under the hypothesis of Theorem~\ref{thm:main}, it holds that
$$
\lim_{n\to\infty}\bP(x(G_n)\in \rint \cX(S)) = \bP(\omega^* \in \Omega(x^*)). 
$$
\end{Proposition}

We establish below Proposition~\ref{prop:convofprobforcv}.
Define an affine transformation $T_n: \R^q \to \R^q$ as follows:
$$x\in \R^q \mapsto T_n(x):= \sqrt{n}(x - x^*) + x^*,$$
which has the inverse given by
$$
\omega\in \R^q \mapsto T^{-1}_n(\omega):=\frac{1}{\sqrt{n}}(\omega - x^*) + x^*.
$$
For each $n\in \mathbb{N}$, we define  
\begin{equation}\label{eq:defsetomegan}
\Omega_n(x^*):= \left\{ T_n(x) \mid x \in \rint \cX(S) \right \}.
\end{equation} 
It should be clear from~\eqref{eq:defomegan} that
$$
x(G_n)\in \rint \cX(S) \, \Longleftrightarrow \, \omega(G_n) = T_n(x(G_n))\in \Omega_n(x^*).
$$
Using~\eqref{eq:hrepfep}, we can characterize the set $\Omega_n(x^*)$ as follows:

\begin{Lemma}\label{lem:Omegan}
It holds that
\begin{equation*}
\Omega_n(x^*) =  
\left\{ \omega\in H_0 \mid  v_\ell^\top \omega > - (\sqrt{n} - 1) v_\ell^\top x^* \mbox{ for } 1\leq \ell \leq k \right\}. 
\end{equation*}
\end{Lemma}

\noindent
{\em Proof.\,}    
From its definition~\eqref{eq:defsetomegan}, we can re-write the set $\Omega_n(x^*)$ as
$$
\Omega_n(x^*) = \left \{ \omega\in \R^q \mid T^{-1}_n(\omega) \in \rint \cX(S) \right \}.
$$
Then, using~\eqref{eq:hrepfep}, we have that $\omega\in \Omega_n(x^*)$ if and only if 
\begin{equation}\label{eq:omegainclusion}
T_n^{-1}(\omega) = \frac{1}{\sqrt{n}}(\omega - x^*) + x^*\in H_0,
\end{equation} 
and 
\begin{equation}\label{eq:omegainequality}
v^\top_\ell T_n^{-1}(\omega) > 0, \quad \mbox{for all } \ell = 1,\ldots, k.
\end{equation}

We show that~\eqref{eq:omegainclusion} holds if and only if $\omega\in H_0$. 
Since $x^*\in H_0$ and since $H_0$ is a hyperplane, we have that $T_n^{-1}(\omega)\in H_0$ if and only if 
$\bfo^\top (\omega - x^*) = 0$, which is equivalent to the condition that $\omega\in H_0$. 

Next, for~\eqref{eq:omegainequality}, we write 
$$
 v^\top_\ell T_n^{-1}(\omega) = 
\frac{1}{\sqrt{n}} v^\top_\ell \omega + \left (1 - \frac{1}{\sqrt{n}} \right )v^\top_\ell x^*.
$$
Thus, 
$$
v^\top_\ell T_n^{-1}(\omega) > 0 \quad \Longleftrightarrow \quad  v_\ell^\top \omega > - (\sqrt{n} - 1) v_\ell^\top x^*.
$$
This completes the proof. \hfill{$\qed$}

We next establish the following result:

\begin{Lemma}\label{lem:l2norm}
There exists a constant $\delta > 0$ such that for any $n \in \mathbb{N}$ and for any $\omega\in \Omega(x^*) - \Omega_n(x^*)$,  
$$
\|\omega\| \geq (\sqrt{n} - 1) \delta.
$$
\end{Lemma}

\noindent
{\em Proof.} We first show that the vectors $v_\ell$, for $\ell = 1,\ldots,k$ span $\R^q$. Suppose, to the contrary, that the vectors $v_\ell$'s do not span $\R^q$; then, there exists a nonzero vector $x$ such that $v^\top_\ell x = 0$ for all $\ell = 1,\ldots, k$. But then, $v^\top_\ell (-x) = 0$ for all~$\ell$. It follows from the half-space representation~\eqref{eq:halfrep} of $\co(S)$ that both $x$ and $-x$ belong to $\co(S)$. However, using the vertex-representation~\eqref{eq:conevertexrep}, we have that both $x$ and $-x$ can be represented as linear combinations of the $z_j$'s with nonnegative coefficients. Since all the entries of the $z_j$'s are nonnegative, $x$ can only be zero, which is a contradiction.   

Next, note that by~\eqref{eq:defcH}, if $H_\ell \in \cH(x^*)$, then $v^\top_\ell x^* = 0$. Thus, by the definition~\eqref{eq:defOmega} of $\Omega(x^*)$ and by Lemma~\ref{lem:Omegan}, we have that 
\begin{equation*}
\Omega(x^*) - \Omega_n(x^*) = \left \{ \omega \in \Omega(x^*) \mid 
v^\top_\ell \omega \leq - (\sqrt{n} - 1) v^\top_\ell x^* \mbox{ for some } \ell \mbox{ s.t. } H_\ell \not\in \cH(x^*) \right \}.
\end{equation*}
Using the fact that the $v_\ell$'s span $\R^q$, we have that the set $\cH -\cH(x^*)$ is nonempty (because otherwise, $v_\ell^\top x^* = 0$ for all $\ell = 1,\ldots, k$, which can hold if and only if $x^* = 0$).  
We then define 
$$\delta := \min\{v_{\ell}^\top x^* \mid H_\ell \notin \cH(x^*)\} > 0.$$
It follows that if $\omega \in \Omega(x^*) - \Omega_n(x^*)$, then there must exist an $H_\ell\notin\cH(x^*)$ such that 
$$
v^\top_\ell \omega \leq  - (\sqrt{n} - 1) v^\top_{\ell} x^* \leq - (\sqrt{n} - 1)\delta.
$$

Finally, using Cauchy-Schwarz inequality and the fact that $v_{\ell}$ is a unit vector, we have that 
$$
\|\omega\| \geq \frac{|v_{\ell}^\top\omega|}{\|v_{\ell}\|} \geq (\sqrt{n} - 1)\delta. 
$$
This completes the proof. \hfill{$\qed$}

With the lemmas above, we are now in the position to prove Proposition~\ref{prop:convofprobforcv}. 

\vspace{.1cm}
\noindent
{\em Proof of Proposition~\ref{prop:convofprobforcv}. \,} We have that
\begin{align}
& |\bP(\omega(G_n) \in \Omega_n(x^*)) - \bP(\omega^* \in \Omega(x^*))| \leq \notag \\
& \qquad\, |\bP(\omega(G_n) \in \Omega_n(x^*)) - \bP(\omega(G_n) \in \Omega(x^*))| \label{firstterm} \\ 
& \quad + |\bP(\omega(G_n) \in \Omega(x^*)) - \bP(\omega^* \in \Omega(x^*))| \label{secondterm} 
\end{align}
We show below that the two terms~\eqref{firstterm} and~\eqref{secondterm} converge to $0$ as $n\to\infty$.

\xc{Proof that~\eqref{firstterm} vanishes.} 
First, we write that
\begin{equation}\label{eq:diffomega}
\bP(\omega(G_n) \in \Omega(x^*)) - \bP(\omega(G_n) \in \Omega_n(x^*))
\\ = \mathbf{P}(\omega(G_n) \in \Omega(x^*)-\Omega_n(x^*)).
\end{equation}
By Lemma~\ref{lem:l2norm}, if $\omega(G_n)\in \Omega(x^*) -\Omega_n(x^*)$, then 
$$
\|\omega(G_n)\| \geq (\sqrt{n} - 1) \delta, 
$$
for some $\delta> 0$ and hence, by triangle inequality, we have that
$$
\|\omega(G_n) - x^*\| \geq (\sqrt{n} - 1)\delta - \|x^*\|.
$$
The above arguments then imply that
\begin{equation}\label{eq:chebpre}
\mathbf{P}(\omega(G_n) \in \Omega(x^*)-\Omega_n(x^*)) \\ \leq \bP(\|\omega(G_n) - x^*\| \geq (\sqrt{n} - 1)\delta - \|x^*\|).
\end{equation}

We will now use the Chebyshev's inequality to bound the expression on the right hand side of~\eqref{eq:chebpre}. To this end, we recall that $nx(G_n)$ is a multinomial random variable with $n$ trials, $q$ events, and $x^*_i$'s the event probabilities. It is well known that $\mathbf{E}[nx(G_n)] = nx^*$ and, moreover,  
its covariance matrix is given by
$$
\operatorname{Cov}(nx(G_n)) = n \Sigma,
$$
where $\Sigma$ is given in Lemma~\ref{lem:limitdist}. 
Using~\eqref{eq:defomegan}, we have that
$$
\omega(G_n) = \frac{1}{\sqrt{n}} (nx(G_n) - nx^*) + x^*,
$$
which implies that $\mathbf{E}[\omega_n] = x^*$ and
\begin{equation}\label{eq:covomegan}
\operatorname{Cov}(\omega(G_n)) = \frac{1}{n} \operatorname{Cov}(nx(G_n)) = \Sigma.
\end{equation}
Now, using Chebyshev's inequality, we obtain that for any $n$ large enough so that $(\sqrt{n} - 1)\delta > \|x^*\|$, 
\begin{equation}\label{eq:chebshev}
\bP(\|\omega(G_n)- x^*\| \geq (\sqrt{n} - 1)\delta - \|x^*\|)  \leq  \frac{\operatorname{tr}(\Sigma)}{\left [(\sqrt{n} - 1)\delta - \|x^*\|\right ]^2}.
\end{equation}
The right hand side of~\eqref{eq:chebshev} vanishes as $n\to\infty$. Consequently, by~\eqref{eq:diffomega},~\eqref{eq:chebpre}, and~\eqref{eq:chebshev}, we have that~\eqref{firstterm} vanishes as $n\to\infty$. 

\xc{Proof that~\eqref{secondterm} vanishes.} This follow from Lemma~\ref{lem:limitdist}; indeed, since the random variable $\omega(G_n)$ converges in distribution to the Gaussian random variable $\omega^*$, we have that
$\lim_{n\to\infty} \bP(\omega(G_n)\in \Omega(x^*)) = \bP(\omega^*\in \Omega(x^*))$. \hfill{$\qed$}

\section{Numerical Study}\label{sec:numericalstudy}
In this section, we conduct numerical studies for three step-graphons, presented in three subsections. For each step-graphon $W_i$, for $i = 1,2,3$, we sampled graphs $\vec{G}_n\sim W_i$ of different orders~$n$. For each $n$, we sample $N = 10,000$ such graphs and compute the empirical probability that $\vec{G}_n$ has a Hamiltonian decomposition (HD).

\subsection{Example 1}

\begin{figure}[t]
    \centering
    \subfloat[\label{sfig1:W1}]{
\begin{tikzpicture}[scale=2.5]

        \def\xOne{0}
        \def\xTwo{1/3}
        \def\xThree{2/3}
        \def\xFour{1}

        \filldraw [fill=black!0, draw=white] (0,1) rectangle (1/3,2/3);
        \filldraw [fill=black!0, draw=white] (\xOne,\xFour) rectangle (\xTwo,\xThree); 
        \filldraw [fill=black!0, draw=white] (\xOne,\xThree) rectangle (\xTwo,\xTwo); 
        \filldraw [fill=black!0, draw=white] (\xTwo,\xFour) rectangle (\xThree,\xThree); 
        \filldraw [fill=black!90, draw=black!90] (\xOne,\xTwo) rectangle (\xTwo,\xOne); 
        \filldraw [fill=black!90, draw=black!90] (\xThree,\xFour) rectangle (\xFour,\xThree); 
        \filldraw [fill=black!50, draw=black!50] (\xTwo,\xThree) rectangle (\xThree,\xTwo); 
        \filldraw [fill=black!35, draw=black!35] (\xTwo,\xTwo) rectangle (\xThree,\xOne); 
        \filldraw [fill=black!35, draw=black!35] (\xThree,\xThree) rectangle (\xFour,\xTwo); 
        \filldraw [fill=black!42, draw=black!42] (\xThree,\xTwo) rectangle (\xFour,\xOne); 

        \draw [draw=black, very thick] (\xOne,\xOne) rectangle (\xFour,\xFour);

        \draw[->, very thick] (\xOne, \xFour) -- (\xOne, \xOne - 0.1) node [left] {$s$};
        \draw[->, very thick] (\xOne, \xFour) -- (\xFour + 0.1, \xFour) node [above] {$t$};

        \node [left] at (\xOne, \xThree) {\small $1/3$};
        \node [left] at (\xOne, \xTwo) {\small $2/3$};

        \node [above] at (\xTwo, \xFour) {\small $1/3$};
        \node [above] at (\xThree, \xFour) {\small $2/3$};

        \node [above left] at (\xOne, \xFour) {\small $0$};   \end{tikzpicture}
}
\qquad
\centering
\subfloat[\label{sfig1:edgepolytope}]{
    \begin{tikzpicture}[scale=1.2]
    \tikzset{every loop/.style={}}
    \node [circle,fill=black,inner sep=0.8pt] (0) at (0,0) {};
    \draw[->] (0,0) -- (1.55,0) node [above] {\footnotesize $x_2$};
    \draw[->] (0,0) -- (0,1.55) node [right] {\footnotesize $x_3$};
    \draw[->] (0,0) -- (-0.9,-0.9) node [left] {\footnotesize $x_1$};

    \node (1) at (1.2, 0) {};
    \node (2) at (0, 1.2) {};
    \node (3) at (-0.66, -0.66) {};
    \node (23) at (-0.33, 0.27) {};

\filldraw[draw=black, fill=gray!0]  (1.center) -- (2.center) -- (3.center) -- (1.center) -- cycle;

\filldraw[draw=black, fill=blue!20]  (1.center) -- (23.center) -- (2.center) -- cycle;

\node [] (xs) at (0.8,1.0) {\small $\cX(S)$};
\node [] (xsl) at (0.8,1.0) {};
\node [] (xst) at (0.7, 0.1) {};

\path[draw,shorten >=3pt,shorten <= 3pt]
    (xsl) edge[-latex] (xst);

\node [circle,fill=red,inner sep=1.0pt,label=below:{\color{red} $x^*$}] (x) at (0.30, 0.15) {};
\end{tikzpicture}
}
\caption{$(a)$ The step-graphon $W_1$. $(b)$ The associated edge polytope $\mathcal{X}(S)$ with concentration vector $x^*$.}
    \label{fig:example1}
\end{figure}
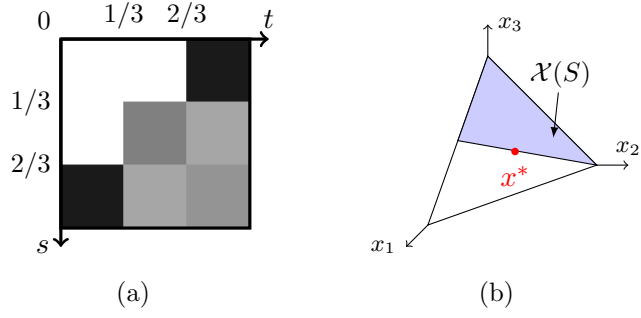

\begin{figure}
    \centering
    \begin{tikzpicture}
        \begin{axis}[
            width=9cm, height=6cm, 
            xlabel={$n$}, 
            ylabel={empirical $\bP(\cE_n)$},
            ymin=-0.05, ymax=1.05, 
            grid=major, 
            minor grid style={dotted}, 
            grid style={dashed, gray!50}, 
            legend pos=south east, 
            xtick={0, 100, 200, 300, 400, 500},
            tick label style={font=\small}, 
        ]
        
        \addplot[
            color=blue, 
            mark=star, 
            dashed, 
            thick
        ] coordinates {
            (20, 0.5218) (60, 0.5232) (100, 0.5294) (200, 0.5128) 
            (300, 0.507) (400, 0.5088) (500, 0.5246)
        };  
        \end{axis}
    \end{tikzpicture}
    \caption{Empirical probability that ${\vec{G}_n}\sim W_1$ has an HD.}
    \label{fig:prob_case1}
\end{figure}
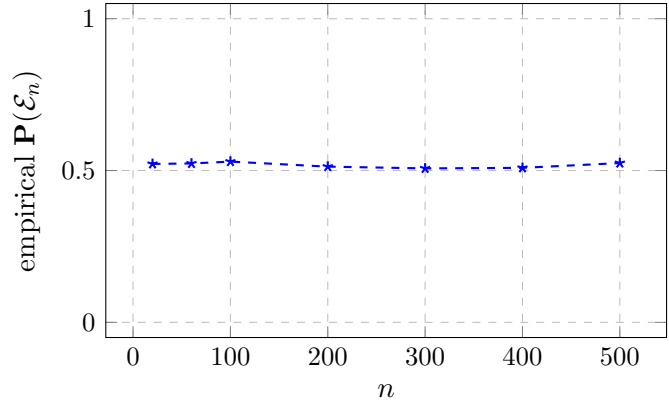

Consider the step-graphon $W_1$ in Figure~\ref{sfig1:W1}, with the concentration vector $x^* = \frac{1}{3}(1, 1, 1)$ on the boundary of the edge polytope shown in Figure~\ref{sfig1:edgepolytope}. 
The associated incidence matrix $Z$ is given by
$$
Z = 
\frac{1}{2}
\begin{bmatrix}
1 & 0 & 0 & 0 \\
0 & 1 & 2 & 0 \\
1 & 1 & 0 & 2 \\
\end{bmatrix}.
$$
The extremal generators of $\cX(S)$ are $z_1$, $z_3$, and $z_4$ (note that $z_2 = (z_3 + z_4)/2$).  
There are three facet-defining hyperplanes $H_\ell$, with the corresponding unit normal vectors $v_\ell$ given by  

\[
\begin{aligned}
   & H_1 = \operatorname{span}\{z_3, z_4\} \quad v_1 = (1, 0, 0), \\
   & H_2 = \operatorname{span}\{z_1, z_3\} \quad  v_2 = \frac{1}{\sqrt{2}}(-1, 0, 1), \\
   & H_3 = \operatorname{span}\{z_1, z_4\} \quad  v_3 = (0, 1, 0).
\end{aligned}
\]
Note that $v^\top_\ell x^* = 0$ if and only if $\ell = 2$. By Corollary~\ref{cor:onehalf}, $\bP(\omega^* \in \Omega(x^*)) = 0.5$. We validate the result in Figure~\ref{fig:prob_case1} where we plotted the empirical probability that $\vec{G}_n\sim W_1$ has an HD for $n \in  \{20, 60, 100, 200, 300, 400, 500\}$.

\subsection{Example 2}
\begin{figure}[h]
\begin{center}
    \begin{tikzpicture}[
        every node/.style={circle, fill=black, inner sep=1pt},
        every path/.style={thick}
    ]
        \node[label=left:$u_1$] (u1) at (0, -0.5) {};
        \node[label=right:$u_2$] (u2) at (0.5, -1) {};
        \node[label=left:$u_3$] (u3) at (-0.5, -1) {};
        \node[label=above:$u_4$] (u4) at (0, 0.1) {};
        \node[label=right:$u_5$] (u5) at (1, -1.5) {};
        \node[label=left:$u_6$] (u6) at (-1, -1.5) {};

        \draw[thick,shorten >=2pt,shorten <=2pt] (u1) -- (u2);
        \draw[thick,shorten >=2pt,shorten <=2pt] (u1) -- (u3);
        \draw[thick,shorten >=2pt,shorten <=2pt] (u1) -- (u4);
        \draw[thick,shorten >=2pt,shorten <=2pt] (u2) -- (u3);
        \draw[thick,shorten >=2pt,shorten <=2pt] (u2) -- (u5);
        \draw[thick,shorten >=2pt,shorten <=2pt] (u3) -- (u6);
        
    \end{tikzpicture}
\end{center}

\caption{Skeleton graph associated with $W_2$ and $W_3$}
\label{fig:skeletongraph}
\end{figure}
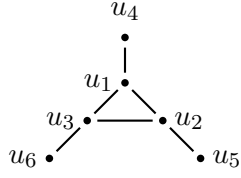

\begin{figure}[h]
    \centering
    \subfloat[\label{sfig1:W2}]{
\begin{tikzpicture}[scale=2.7]

    \filldraw [fill=black!0, draw=black!0] (0, 0.25) rectangle (0.25, 0.125); 
    \filldraw [fill=black!0, draw=black!0] (0.75, 1) rectangle (0.875, 0.75); 

    \filldraw [fill=black!0, draw=black!0] (0, 0.125) rectangle (0.25, 0); 
    \filldraw [fill=black!0, draw=black!0] (0.875, 1) rectangle (1, 0.75); 

    \filldraw [fill=black!0, draw=black!0] (0.25, 0.375) rectangle (0.5, 0.25); 
    \filldraw [fill=black!0, draw=black!0] (0.625, 0.75) rectangle (0.75, 0.5); 

    \filldraw [fill=black!0, draw=black!0] (0.25, 0.125) rectangle (0.5, 0); 
    \filldraw [fill=black!0, draw=black!0] (0.875, 0.75) rectangle (1, 0.5); 

    \filldraw [fill=black!0, draw=black!0] (0.5, 0.375) rectangle (0.625, 0.25); 
    \filldraw [fill=black!0, draw=black!0] (0.625, 0.5) rectangle (0.75, 0.375); 

    \filldraw [fill=black!0, draw=black!0] (0.5, 0.25) rectangle (0.625, 0.125); 
    \filldraw [fill=black!0, draw=black!0] (0.75, 0.5) rectangle (0.875, 0.375); 

    \filldraw [fill=black!0, draw=black!0] (0.75, 0.375) rectangle (0.875, 0.25); 
    \filldraw [fill=black!0, draw=black!0] (0.625, 0.25) rectangle (0.75, 0.125); 

    \filldraw [fill=black!0, draw=black!0] (0.75, 0.125) rectangle (0.875, 0); 
    \filldraw [fill=black!0, draw=black!0] (0.875, 0.25) rectangle (1, 0.125); 

    \filldraw [fill=black!0, draw=black!0] (0,1) rectangle (0.25, 0.75); 

    \filldraw [fill=black!0, draw=black!0] (0.25, 0.75) rectangle (0.5, 0.5);

    \filldraw [fill=black!0, draw=black!0] (0.5, 0.5) rectangle (0.625, 0.375);

    \filldraw [fill=black!0, draw=black!0] (0.625, 0.375) rectangle (0.75, 0.25);

    \filldraw [fill=black!0, draw=black!0] (0.75, 0.25) rectangle (0.875, 0.125);

    \filldraw [fill=black!0, draw=black!0] (0.875, 0.125) rectangle (1, 0);


    \filldraw [fill=black!70, draw=black!70] (0, 0.75) rectangle (0.25, 0.5); 
    \filldraw [fill=black!70, draw=black!70] (0.25, 1) rectangle (0.5, 0.75); 

    \filldraw [fill=black!70, draw=black!70] (0, 0.5) rectangle (0.25, 0.375); 
    \filldraw [fill=black!70, draw=black!70] (0.5, 1) rectangle (0.625, 0.75); 

    \filldraw [fill=black!70, draw=black!70] (0, 0.375) rectangle (0.25, 0.25); 
    \filldraw [fill=black!70, draw=black!70] (0.625, 1) rectangle (0.75, 0.75); 

    \filldraw [fill=black!70, draw=black!70] (0.25, 0.5) rectangle (0.5, 0.375); 
    \filldraw [fill=black!70, draw=black!70] (0.5, 0.75) rectangle (0.625, 0.5); 

    \filldraw [fill=black!70, draw=black!70] (0.25, 0.25) rectangle (0.5, 0.125); 
    \filldraw [fill=black!70, draw=black!70] (0.75, 0.75) rectangle (0.875, 0.5); 

    \filldraw [fill=black!70, draw=black!70] (0.5, 0.125) rectangle (0.625, 0); 
    \filldraw [fill=black!70, draw=black!70] (0.875, 0.5) rectangle (1, 0.375); 

        \draw [draw=black, very thick] (0,0) rectangle (1,1);

        \draw[->, very thick] (0, 1) -- (0, -0.1) node [left] {$s$};
        \draw[->, very thick] (0, 1) -- (1.1, 1) node [above] {$t$};

        \node [left] at (0, 0.75) {\scriptsize$2/8$};
        \node [left] at (0, 0.5) {\scriptsize$4/8$};
        \node [left] at (0, 0.375) {\scriptsize$5/8$};
        \node [left] at (0, 0.25) {\scriptsize$6/8$};
        \node [left] at (0, 0.125) {\scriptsize$7/8$};

        \node [above] at (0.25, 1) {$\frac{2}{8}$};
        \node [above] at (0.5, 1) {$\frac{4}{8}$};
        \node [above] at (0.625, 1) {$\frac{5}{8}$};
        \node [above] at (0.75, 1) {$\frac{6}{8}$};
        \node [above] at (0.875, 1) {$\frac{7}{8}$};

        \node [left] at (0, 1.05) {$0$}; 

    \end{tikzpicture}
}
\centering
\subfloat[\label{sfig1:w3}]{
   \begin{tikzpicture}[scale=2.7]

        \filldraw [fill=black!0, draw=black!0] (0,1) rectangle (0.1667, 0.8333); 

        \filldraw [fill=black!0, draw=black!0] (0.1667, 0.8333) rectangle (0.3333, 0.6667);

        \filldraw [fill=black!0, draw=black!0] (0.3333, 0.6667) rectangle (0.5, 0.5);

        \filldraw [fill=black!0, draw=black!0] (0.5, 0.5) rectangle (0.6667, 0.3333);

        \filldraw [fill=black!0, draw=black!0] (0.6667, 0.3333) rectangle (0.8333, 0.1667);

        \filldraw [fill=black!0, draw=black!0] (0.8333, 0.1667) rectangle (1, 0);


        \filldraw [fill=black!0, draw=black!0] (0.1667, 0.5) rectangle (0.3333, 0.3333); 
        \filldraw [fill=black!0, draw=black!0] (0.5, 0.8333) rectangle (0.6667, 0.6667); 

        \filldraw [fill=black!0, draw=black!0] (0.1667, 0.1667) rectangle (0.3333, 0); 
        \filldraw [fill=black!0, draw=black!0] (0.8333, 0.8333) rectangle (1, 0.6667); 

        \filldraw [fill=black!0, draw=black!0] (0.3333, 0.5) rectangle (0.5, 0.3333); 
        \filldraw [fill=black!0, draw=black!0] (0.5, 0.6667) rectangle (0.6667, 0.5); 

        \filldraw [fill=black!0, draw=black!0] (0.3333, 0.3333) rectangle (0.5, 0.1667); 
        \filldraw [fill=black!0, draw=black!0] (0.6667, 0.5) rectangle (0.8333, 0.3333); 

        \filldraw [fill=black!0, draw=black!0] (0.6667, 1) rectangle (0.8333, 0.8333); 
        \filldraw [fill=black!0, draw=black!0] (0, 0.3333) rectangle (0.1667, 0.1667); 

        \filldraw [fill=black!0, draw=black!0] (0.8333, 0.3333) rectangle (1, 0.1667); 
        \filldraw [fill=black!0, draw=black!0] (0.5, 0.1667) rectangle (0.6667, 0); 


        \filldraw [fill=black!70, draw=black!70] (0, 0.8333) rectangle (0.1667, 0.6667); 
        \filldraw [fill=black!70, draw=black!70] (0.1667, 1) rectangle (0.3333, 0.8333); 

        \filldraw [fill=black!70, draw=black!70] (0, 0.6667) rectangle (0.1667, 0.5); 
        \filldraw [fill=black!70, draw=black!70] (0.3333, 1) rectangle (0.5, 0.8333); 

        \filldraw [fill=black!70, draw=black!70] (0, 0.5) rectangle (0.1667, 0.3333); 
        \filldraw [fill=black!70, draw=black!70] (0.5, 1) rectangle (0.6667, 0.8333); 

        \filldraw [fill=black!70, draw=black!70] (0.1667, 0.6667) rectangle (0.3333, 0.5); 
        \filldraw [fill=black!70, draw=black!70] (0.3333, 0.8333) rectangle (0.5, 0.6667); 

        \filldraw [fill=black!70, draw=black!70] (0.1667, 0.3333) rectangle (0.3333, 0.1667); 
        \filldraw [fill=black!70, draw=black!70] (0.6667, 0.8333) rectangle (0.8333, 0.6667); 

        \filldraw [fill=black!70, draw=black!70] (0.3333, 0.1667) rectangle (0.5, 0); 
        \filldraw [fill=black!70, draw=black!70] (0.8333, 0.6667) rectangle (1, 0.5); 

        \node [left] at (0, 5/6) {\scriptsize$1/6$};
        \node [left] at (0, 4/6) {\scriptsize$2/6$};
        \node [left] at (0, 3/6) {\scriptsize$3/6$};
        \node [left] at (0, 2/6) {\scriptsize$4/6$};
        \node [left] at (0, 1/6) {\scriptsize$5/6$};

        \node [above] at (1/6, 1) {$\frac{1}{6}$};
        \node [above] at (2/6, 1) {$\frac{2}{6}$};
        \node [above] at (3/6, 1) {$\frac{3}{6}$};
        \node [above] at (4/6, 1) {$\frac{4}{6}$};
        \node [above] at (5/6, 1) {$\frac{5}{6}$};

        \node [left] at (0, 1.05) {$0$}; 

        \draw [draw=black, very thick] (0,0) rectangle (1,1);

        \draw[->, very thick] (0, 1) -- (0, -0.1) node [left] {$s$};
        \draw[->, very thick] (0, 1) -- (1.1, 1) node [above] {$t$};

    \end{tikzpicture}
}
\caption{$(a)$ Step-graphon $W_2$. $(b)$ Step-graphon $W_3$.}
    \label{fig:graphone2e3}
\end{figure}
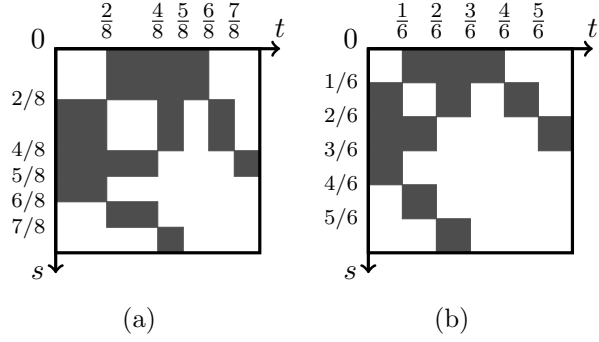

Consider the step-graphon $W_2$ in Figure~\ref{sfig1:W2} with the concentration vector $x^* =\frac{1}{8}(2, 2, 1, 1, 1, 1)$. 
The associated skeleton graph is given in Figure~\ref{fig:skeletongraph}.

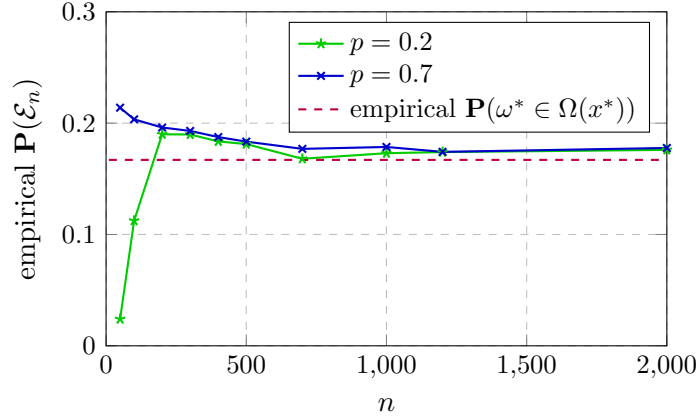
\begin{figure}[h]
    \centering
    \begin{tikzpicture}
    \begin{axis}[
        width=9cm, height=6cm, 
        xlabel={$n$},
        ylabel={empirical $\bP(\cE_n)$},
        xmin=0, xmax=2000, xtick={0,500,1000, 1500, 2000},
        ymin=0, ymax=0.30,
        grid=major,
        minor grid style=dotted,
        major grid style=dashed,
        legend pos=north east,
        legend style={draw=black, fill=white, font=\small, legend cell align=left},
        tick label style={font=\small}
    ]
     \addplot[thick, color=green!80!black, mark=star, ]
    coordinates {
        (50, 0.024) (100, 0.1124) (200, 0.19) (300, 0.1898) (400, 0.1836)
        (500, 0.1812) (700, 0.168) (1000, 0.173) (1200, 0.174) (2000, 0.176)
    };
    \addlegendentry{$p = 0.2$}

    \addplot[thick, color=blue!80!black, mark=x,] coordinates {
        (50, 0.2139) (100, 0.2034) (200, 0.1961) (300, 0.193) (400, 0.1875) (500, 0.1834) (700, 0.1769) (1000, 0.1786) (1200, 0.1742) (2000, 0.1777)
    };
    \addlegendentry{$p = 0.7$}
    
    \addplot[thick, color=purple, dashed] coordinates {
        (10, 0.16699) (50, 0.16699) (70, 0.16699) (100, 0.16699) 
        (125, 0.16699) (150, 0.16699) (175, 0.16699) (200, 0.16699) 
        (300, 0.16699) (400, 0.16699) (500, 0.16699) (700, 0.16699) 
        (1000, 0.16699) (1200, 0.16699) (2000, 0.16699)
    };
    \addlegendentry{empirical $\bP(\omega^*\in \Omega(x^*))$}
    \end{axis}
\end{tikzpicture}
    \caption{Empirical probability that ${\vec{G}_n}\sim W_2$ has an HD and empirical probability that $\omega^*\in \Omega(x^*)$ (with 200,000 samples), which is approximately 0.16699.}
    \label{fig:epw2}
\end{figure}

The associated incidence matrix $Z$ is given by:
\[
Z = 
\frac{1}{2}
\begin{bmatrix}
1 & 1 & 1 & 0 & 0 & 0 \\
1 & 0 & 0 & 1 & 1 & 0 \\
0 & 1 & 0 & 1 & 0 & 1 \\
0 & 0 & 1 & 0 & 0 & 0 \\
0 & 0 & 0 & 0 & 1 & 0 \\
0 & 0 & 0 & 0 & 0 & 1
\end{bmatrix}
\]
In this case, all the columns of $Z$ are extremal generators~\cite{belabbas2021h}. 
There are six facet-defining hyperplanes $H_\ell$ (each of which is spanned by all but one $z_j$'s), with the corresponding $v_\ell$ given by  
\[
\begin{aligned}
   & H_1 = \operatorname{span}\{z_1, z_2, z_3, z_4, z_5\} \quad v_1 = (0, 0, 0, 0, 0, 1), \\
   & H_2 = \operatorname{span}\{z_1, z_2, z_3, z_4, z_6\} \quad  v_2 = (0, 0, 0, 0, 1, 0), \\
   & H_3 = \operatorname{span}\{z_1, z_2, z_3, z_5, z_6\} \quad  v_3 = \frac{1}{\sqrt{6}}(-1, 1, 1, 1, -1, -1),\\
    & H_4 = \operatorname{span}\{z_1, z_2, z_4, z_5, z_6\} \quad v_4 = (0, 0, 0, 1, 0, 0), \\
   & H_5 = \operatorname{span}\{z_1, z_3, z_4, z_5, z_6\} \quad  v_5 = \frac{1}{\sqrt{6}}(1, -1, 1, -1, 1, -1), \\
   & H_6 = \operatorname{span}\{z_2, z_3, z_4, z_5, z_6\} \quad  v_6 = \frac{1}{\sqrt{6}}(1, 1, -1, -1, -1, 1).\\
\end{aligned}
\]
Observe that $v_\ell^{\top}x^* = 0$ for $\ell = 3, 5$. In Figure~\ref{fig:epw2}, we plot the empirical probability of the event $\vec{G}_n$ has an HD for $n \in  \{50, 100, 200, 300, 400, 500, 700, 1000, 1200, 2000\}$ and compare the results with the empirical probability of the event $\omega^*\in \Omega(x^*)$ where we draw $200,000$ {\em i.i.d.} Gaussian random variables from $N(x^*,\Sigma)$.    

\subsection{Example 3}
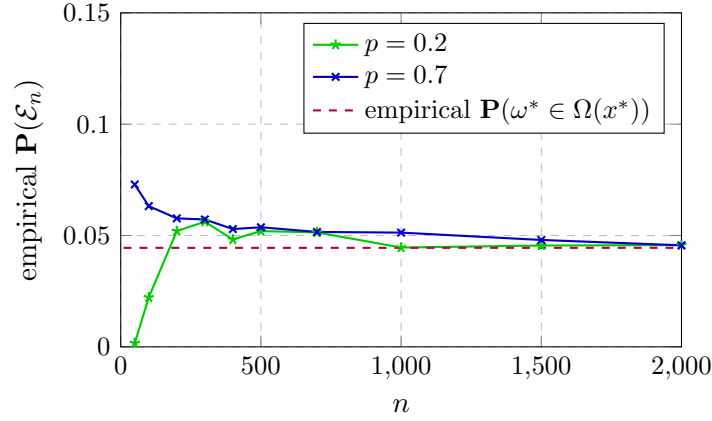
\begin{figure}[h]
    \centering
    \begin{tikzpicture}
    \begin{axis}[
        width=9cm, height=6cm, 
        xlabel={$n$},
        ylabel={empirical $\bP(\cE_n)$},
        xmin=0, xmax=2000, xtick={0,500,1000, 1500, 2000},
        ymin=0, ymax=0.15,
        grid=major,
        minor grid style=dotted,
        major grid style=dashed,
        legend pos=north east,
        legend style={draw=black, fill=white, font=\small, legend cell align=left},
        tick label style={font=\small},
        scaled y ticks=false, 
    yticklabel style={/pgf/number format/fixed} 
    ]
    
    \addplot[thick, color=green!80!black, mark=star,] coordinates {
        (50, 0.0016) (100, 0.0222) (200, 0.052) (300, 0.0562) (400, 0.0482) (500, 0.052) (700, 0.0514) (1000, 0.0446) (1500, 0.0455) (2000, 0.0457)
    };
    \addlegendentry{$p = 0.2$}
    
    \addplot[thick, color=blue!80!black, mark=x] coordinates {
        (50, 0.0729) (100, 0.0632) (200, 0.0577) (300, 0.0572) (400, 0.0529) (500, 0.0537) (700, 0.0516) (1000, 0.0513) (1500, 0.0480) (2000, 0.0456)
    };
    \addlegendentry{$p = 0.7$}
    \addplot[thick, color=purple, dashed] coordinates {
        (10, 0.04446) (50, 0.04446) (70, 0.04446) (100, 0.04446) 
        (125, 0.04446) (150, 0.04446) (175, 0.04446) (200, 0.04446) 
        (300, 0.04446) (400, 0.04446) (500, 0.04446) (700, 0.04446) 
        (1000, 0.04446) (1500, 0.04446) (2000, 0.04446)
    };
    \addlegendentry{empirical $\bP(\omega^*\in \Omega(x^*))$}
    \end{axis}
\end{tikzpicture}
    \caption{Empirical probability that ${\vec{G}_n}\sim W_3$ has an HD and empirical probability that $\omega^*\in \Omega(x^*)$ (with 200,000 samples), which is approximately 0.04446.}
    \label{fig:epw3}
\end{figure}

Finally, we consider the step-graphon $W_3$ shown in Figure~\ref{sfig1:w3} with the concentration vector $x^* =\frac{1}{6}(1, 1, 1, 1, 1, 1)$. Note that $W_2$ and $W_3$ share the same skeleton graph, the incidence matrix $Z$, and the collection of facet-defining hyperplanes. However, in this case, we have that $v_\ell^{\top}x^* = 0$ for $\ell = 3, 5, 6$, so $\mathcal{H}(x^*) = \{H_3,H_5,H_6\}$ (compared to the previous case where $\mathcal{H}(x^*) = \{H_3,H_5\}$). 
Similar to what has been done for the previous case, we plot in Figure~\ref{fig:epw3} the empirical probability of the event that $\vec{G}_n$ has an HD for $n \in  \{50, 100, 200, 300, 400, 500, 700, 1000, 1500, 2000\}$ and compare the results with the empirical probability of the event $\omega^*\in \Omega(x^*)$.    

\section{Conclusion}
In this paper, we draw random graphs ${\vec{G}_n}$ from a step-graphon $W$ and investigate the probability that ${\vec{G}_n}$ has 
a Hamiltonian decomposition as $n\to\infty$. It has been shown in the earlier works that for almost all step-graphons $W$, this probability converges to either zero or one, depending on whether the associated edge polytope $\cX(S)$ is full rank and whether the concentration vector $x^*$ is contained in (the relative interior of) $\cX(S)$ (see Theorem~\ref{thm:zeroone}). 
The case that had been left out was the one such that $\cX(S)$ has full rank and $x^*\in \partial\cX(S)$, we termed such case the {\em residual case}. 
The main contribution of this paper is to show that for this case the aforementioned probability still converges and, moreover, to provide an explicit expression of the limit. This result has been formulated in Theorem~\ref{thm:main}.   

\printbibliography

\end{document}